\documentclass[%
 reprint,
 amsmath,amssymb,
 aps,
prb,
]{revtex4-2}

\usepackage{amsmath}
\usepackage{amsfonts}
\usepackage{bm}
\usepackage{braket}
\usepackage{caption}
\usepackage{float}
\usepackage{gensymb}
\usepackage{graphicx}
\usepackage[latin1,utf8]{inputenc}
\usepackage{mathtools}
\usepackage{orcidlink}
\usepackage{subcaption}
\usepackage{upgreek}

\usepackage{lipsum}

\usepackage{lineno}

\DeclareMathSymbol{\shortminus}{\mathbin}{AMSa}{"39}

 \usepackage{fancyhdr}
 \fancypagestyle{firstpage}{
 \fancyhf{}\lfoot{DISTRIBUTION A. Approved for public release: distribution unlimited. AFRL-2025-XXXX}
 }
 \fancypagestyle{plain}{%
   \fancyhf{} 
   \fancyfoot[C]{DISTRIBUTION A. Approved for public release: distribution unlimited.}
   
 }
 
\begin{document}

\title{Experimental Measurement of Enhanced Group Delay Silicon Photonic Waveguides Indicative of the Frozen Mode Regime Around the Stationary Inflection Point}

\author{Nathaniel~Furman$^{1}$\orcidlink{0000-0001-7896-2929},
        Albert~Herrero-Parareda$^{1}$\orcidlink{0000-0002-8501-5775},
        Anthony~Rapp$^{3,4}$\orcidlink{0000-0001-8871-7361},
        Ilya~Vitebski$y^{2}$\orcidlink{0000-0001-8375-2088}
}

        \author{Ricky~Gibson$^{2}$\orcidlink{0000-0002-2567-6707}}
        \email{ricky.gibson.2@us.af.mil}
\author{
        Bradley~J.~Thompson$^{2}$\orcidlink{0000-0002-7416-2112},
        Dean~P.~Brown$^{3,5}$,
        Robert~Bedford$^{3}$\orcidlink{0000-0002-0457-0081}
}

\author{Filippo~Capolino$^{1}$\orcidlink{0000-0003-0758-6182}}
\email{f.capolino@uci.edu}

\affiliation{
\mbox{$^{1}$Department
of Electrical Engineering and Computer Science, University of California, Irvine, CA, 92697 USA}
\\
\mbox{$^{2}$Air Force Research Laboratory, Sensors Directorate, Wright-Patterson Air Force Base, Ohio, 45433 USA}
\\
\mbox{$^{3}$ Air Force Research Laboratory, Materials and Manufacturing Directorate, Wright-Patterson Air Force Base, Ohio, 45433 USA}
\\
\mbox{$^{4}$Department of Electro-Optics and Photonics, University of Dayton, Ohio, 45469 USA}
\\
\mbox{$^{5}$UES, a BlueHalo company, Ohio, 45432 USA}
}

\begin{abstract}
The dispersion engineering of periodic silicon photonic waveguides presents opportunities for significant group delay enhancement compared to uniform waveguides of comparable length. We describe the spectral response characteristics for measured devices and compare their properties to modeled data. These waveguides support the frozen mode regime (FMR) around near infrared wavelengths and are expected to show enhanced group delays around the FMR resonances. Measurements of fabricated devices provide evidence for enhanced delays and spectral properties associated with the FMR. We study how perturbations to the waveguide model impact agreement with measurements and its meaning for these devices operating in the FMR.
\end{abstract}

\maketitle

\section{Introduction}
\label{ch:Intro}

Periodic waveguides represent a wide class of structures used for a variety of applications, and in particular we consider applications in dispersion engineering \cite{schulzDispersionEngineeredSlow2010,sakoda_enhanced_1999,figotin_gigantic_2005,ballatoFrozenLightPeriodic2005,sukhorukovSlowlightDispersionCoupled2008,burr_degenerate_2013}. Both in low frequency microstrip geometries (gigahertz range) and high frequency photonic platforms (hundreds of terahertz range), periodic devices can exhibit properties generally associated with exotic materials rather than with widely used fabrication processes \cite{figotin_oblique_2003,altug_experimental_2005,maasExperimentalRealizationEpsilonnearzero2013,stephansonFrozenModesCoupled2008,campioneComplexModesZero2011,othman_experimental_2017}. Of increasing recent interest is dispersion engineering waveguides for dispersion flat-band conditions where the group velocity is near zero. These regions are associated with significantly large group delays and further an anomalous scaling of group delay with waveguide length. This slow down of light around particular dispersion conditions, commonly called the frozen mode regime~(FMR), has potential applications in low-threshold lasing, optical switching and delay lines, nonlinear optics, sensors, and more \cite{kimNanobeamPhotonicBandedge2011,paulFrozenModeCoupled2021,gutman_slow_2012,burrExperimentalVerificationDegenerate2016,ramezani_unidirectional_2014,dowlingPhotonicBandEdge1994,scaloraOpticalLimitingSwitching1994,gutmanFrozenBroadbandSlow2012,nadaDesignModifiedCoupled2023,nada_theory_2017,nada_giant_2018,baba_slowlight_2014}.

We focus on a particular type of dispersion-engineered periodic structure resultant from the coalescence of multiple Floquet-Bloch eigenmodes in the infinitely-periodic extent. These dispersion degeneracies are referred to as exceptional points of degeneracy~(EPDs) and occur when at least two eigenstates (in both their eigenvalues and eigenvectors) of a waveguide coalesce at a singular point in the parameter space. A subset of EPDs can have orders two, three, or four depending on the number of modes coalescing at a particular frequency. Degeneracies in waveguides of orders two, three, and four are referred to as the regular band edge~(RBE), the stationary inflection point~(SIP), and the degenerate band edge~(DBE), respectively \cite{figotin_frozen_2006,figotin_oblique_2003,figotin_gigantic_2005,figotin_slow_2006}.

The waveguide presented here is designed to exhibit the SIP in its Floquet-Bloch dispersion relation. The dispersion near the SIP is well approximated by a cubic relationship $(\omega - \omega_0) \propto (k - k_0)^3$ where $\omega$ is the angular frequency and $k$ is the wavenumber. For resonances around the SIP frequency, the group delay scales asymptotically as $N^3$ for large $N$, where $N$ represents the number of waveguide unit cells in finite-length structures \cite{herrero-parareda_frozen_2022,furmanFrozenModeRegime2023,nadaDesignModifiedCoupled2023}. One key difference between the SIP and the RBE/DBE is the lack of a bandgap at frequencies adjacent to the SIP. The RBE and DBE tend to have their wavenumber degeneracy at a Brillouin Zone~(BZ) boundary (i.e., $kd/\pi$ is zero or one where $d$ represents the unit cell period) whereas the SIP wavenumber is within the extents of the BZ. The precise location of the SIP wavenumber inside the BZ is thought to be geometry specific with some ability to tune based on the particular choice of design geometric parameters.

A couple of critical elements needed for a waveguiding structure to exhibit the SIP in its dispersion diagram are three modes in each $+z$ and $\shortminus z$ directions that mix, i.e., they couple in a continuous or periodic fashion. While the general conditions for EPD formation in three-mode waveguides have been detailed in Ref.~\cite{nadaGeneralConditionsRealize2018}, more specifically for the SIP there must be some way to facilitate counter-propagating modes interacting. Some geometries achieve this with coupled microring resonators, while others use photonic crystals or gratings. Waveguides that support more than three modes at a given frequency (in each direction) can also exhibit SIP behavior; yet to the best of our knowledge, that is an outstanding area of research.

This work showcases multiple fabricated silicon photonic waveguide spectral responses and group delays when varying device lengths. We compare the measured results with models to help establish waveguide operation near the SIP and FMR. In Sec.~\ref{ch:Geometries}, we present the periodic device with its parameters, geometric configurations, and other pertinent information. Sections~\ref{ch:Setup}~and~\ref{ch:Results} detail the experimental measurement setup and results. We discuss the results in Sec.~\ref{ch:Discussion}.

\section{Waveguide Geometry and SIP}
\label{ch:Geometries}

\begin{figure*}[!t]
    \centering
        \includegraphics[width=0.98\textwidth]{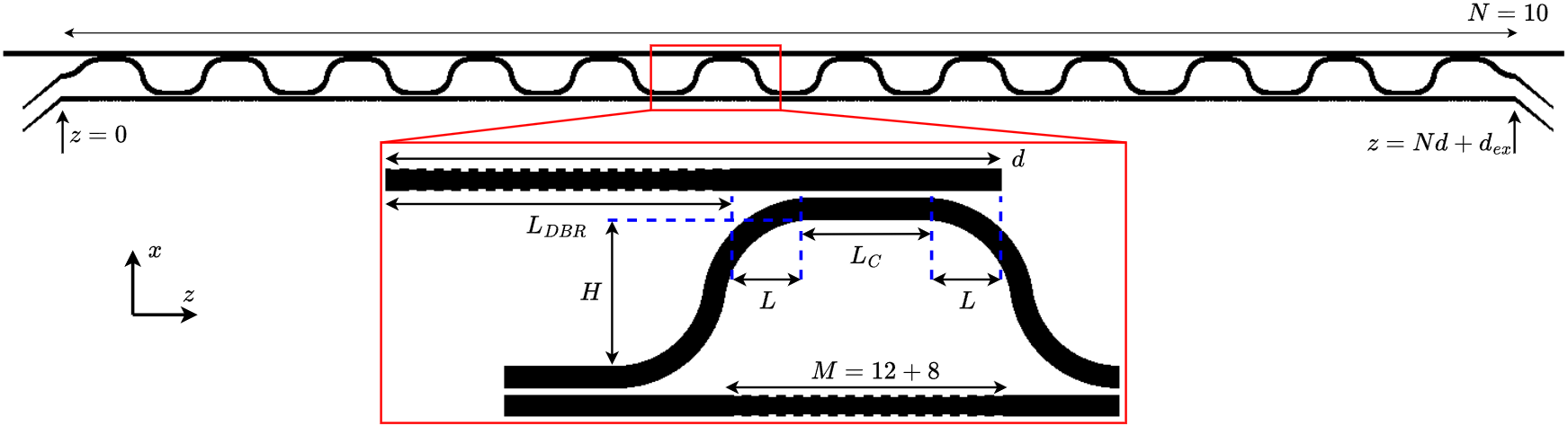}
    \\
    \caption{Finite-length geometry of the 3PD waveguide for $N=10$ unit cells with unit cell geometry enlarged and marked with parameters given in Table~\ref{tab:ParameterValues-3PD}. The DBR in the 3PD unit cell has $M = 12 + 8$ segments where $12$ are fully perturbed and four on each side have gradually decreasing inner waveguide widths (apodized grating). The 3PD has an additional termination geometry at each end working to separate the coupled waveguides and reduce the reflection at the waveguide boundaries. The continuous straight paths on the top waveguide represent the input and output waveguides.}
    \label{fig:Geometry-3PD}
\end{figure*}

The waveguide presented here is referred to as the 3PD geometry. The 3PD, with its name derived from a three-path waveguide with distributed Bragg reflector, has been discussed with more detail in Refs.~\cite{furmanFrozenModeRegime2023,furmanImpactWaveguideImperfections2023,furmanImpactFabricationDisorder2025}. The unit cell for the waveguide is shown in Fig.~\ref{fig:Geometry-3PD} with its corresponding geometric parameters given in Table~\ref{tab:ParameterValues-3PD}. The 3PD waveguide was designed with a hybrid analytic and simulation approach with full-wave finite element method~(FEM) electromagnetic simulations and was designed to exhibit the SIP in its Floquet-Bloch dispersion relation within the telecommunications C-band. A full electric field description and modeling overview is included in the prior citations. When referring to the waveguide width $w$, the value given is the width at the base, or bottom, of the waveguide. The fabricated waveguides are assumed to have an $\approx 85\degree$ sidewall angle based on measurements from fabricated test structures where sidewall angle was used to test the effective refractive index and group index of ring resonators.

\begin{table*}
    \centering
    \begin{tabular}{l|rrrr|rrrr}
        \textbf{Parameter} & $w\;\mathrm{(nm)}$ & $h\;\mathrm{(nm)}$ & $g\;\mathrm{(nm)}$ & $d\;\mathrm{(\upmu m)}$ & $L\;\mathrm{(\upmu m)}$ & $L_c\;\mathrm{(\upmu m)}$ & $H\;\mathrm{(\upmu m)}$ & $L_{DBR}\;\mathrm{(\upmu m)}$ \\
        \hline
         & 450 & 220 & 150 & 27.0 &  8.492 & 2.728 & 5.597 &  7.3   \\
    \end{tabular}
    \caption{Model parameters for the 3PD waveguide where $w$ is the waveguide width, $h$ is the waveguide height, $g$ is the gap size between coupled waveguides, $d$ is the unit cell period length, and other design-specific parameters are shown in Fig.~\ref{fig:Geometry-3PD} where the values have been rounded to the nearest nanometer. The 3PD design is optimized for $\lambda_{\mathrm{SIP}} = 1553.3\;\mathrm{nm}$.}
    \label{tab:ParameterValues-3PD}
\end{table*}

\begin{figure}
    \centering
        \includegraphics[width=0.48\textwidth]{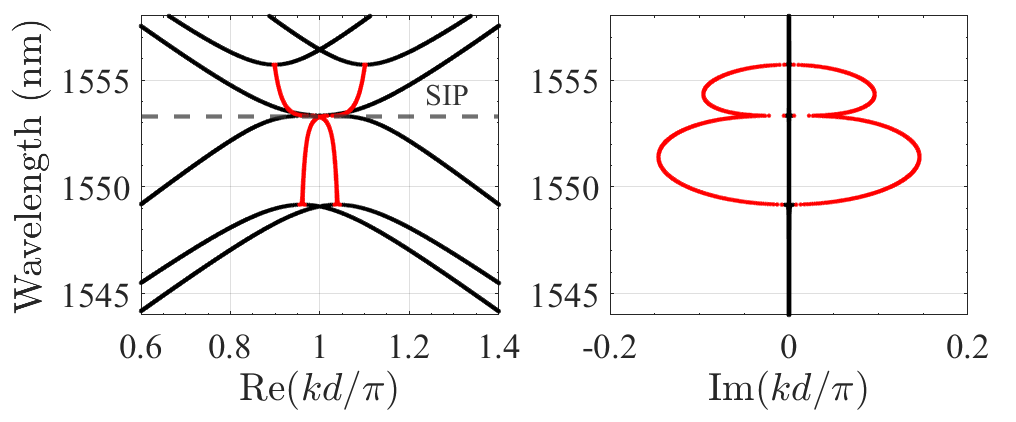}
    \caption{The Floquet-Bloch dispersion diagram for the 3PD waveguide's modes where the SIP wavelength is marked with a horizontal dashed line. Propagating modes are shown in black and evanescent modes are shown in red. At each of the two SIPs, symmetric with respect to the $kd = \pi$ point, all three of the coalescing wavenumbers are real.}
    \label{fig:Dispersion}
\end{figure}

The modal dispersion diagram, calculated by solving the dispersion characteristic equation
\begin{equation}
    D(k,\omega) \equiv \det{[\,\boldsymbol{\underline{T}}_U - \zeta \boldsymbol{\underline{1}}\,]}=0,
    \label{eq:Dispersion-Equation}
\end{equation}
is shown in Fig.~\ref{fig:Dispersion}. In this equation, $\boldsymbol{\underline{T}}_U$ is the unit cell transfer matrix evolving the electric field state vector through a waveguide unit cell with period $d$ and $\zeta \equiv e^{-jkd}$ is the eigenvalue of the system with Bloch wavenumber $k$. $\boldsymbol{\underline{1}}$ represents the $6\mathrm{x}6$ identity matrix. In Fig.~\ref{fig:Dispersion}, black curves represent propagating modes (purely real wavenumber) whereas red curves represent evanescent modes (complex-valued wavenumber).
The diagram shows two SIPs, symmetrically located with respect to the $kd = \pi$ center point. At each SIP, three eigenmodes coalesce (black and red curves) resulting in a real wavenumber. At frequencies near the SIP frequency, we observe other spectral degeneracies. Namely, two RBEs with the closest RBE having hundreds of gigahertz (few nanometer) spectral separation from the SIP. A larger separation is more helpful when finite-length waveguides are excited near the SIP frequency as the waveguide's SIP resonances are better isolated from other EPD resonances. The spectral region between the two RBEs (later on called the ``pseudo bandgap'') has a single mode propagating in each direction.

Using the unit cell models, we calculate the finite-length waveguide transfer function (also called the waveguide spectral response). We cascade the waveguide unit cells $N$ times and include termination, or boundary condition, geometries before the first unit cell and after the last unit cell. Specifically, we calculate the transfer function as
\begin{equation}
    T_f=\frac{E_1^+(z = Nd+d_{ex})}{E_1^+(z = 0)},
    \label{eq:TransferFunc}
\end{equation}
where $E_1^+(z)$ represents the electric field phasor propagating in the $+z$ direction in the primary waveguide used for input and output, $d$ is the unit cell period, and $d_{ex}$ is the total additional length from all termination geometries. Here, $d_{ex} = 2d - L_c$. These terminations represent modifications to the unit-cell geometry and are seen at the waveguide extents in Fig.~\ref{fig:Geometry-3PD}. Using the transfer function phase, we calculate the group delay as
\begin{equation}
    \tau_g=-\frac{\partial \angle T_f(\omega)}{\partial \omega},
    \label{eq:GroupDelay}
\end{equation}
where $\omega=2\pi f=2\pi c_0 / \lambda$ is the angular frequency and $c_0$ is the speed of light in vacuum.

\begin{figure}
    \centering
    \includegraphics[width=0.48\textwidth]{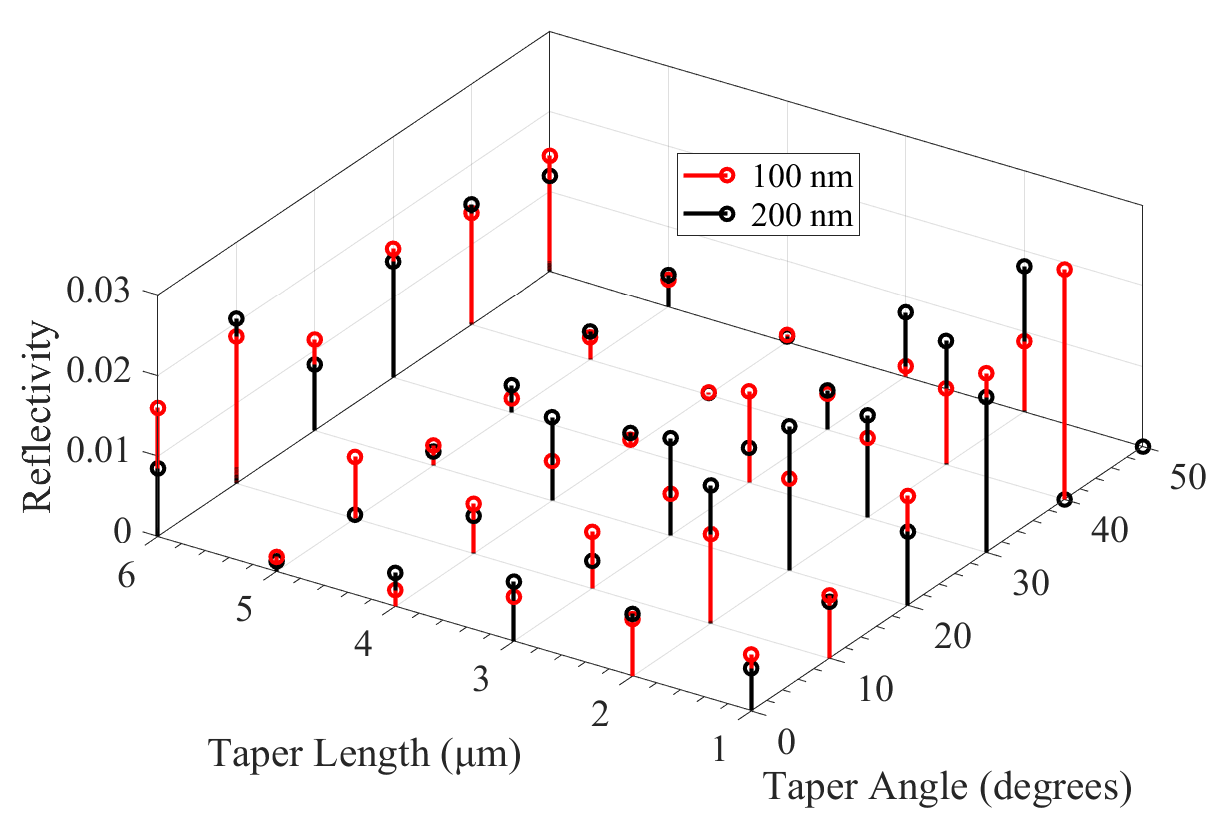}
    \caption{Reflectivity of the termination tapers across a subset of the design space when varying the taper length, angle, and minimum width. The length is defined as the projection of the taper on the $z$-axis, the angle is defined from the $z$-axis, and the minimum width is defined as the width of the waveguide at the end of the taper ($100$ or $200\;\mathrm{nm}$). The geometric parameters of the tapers were chosen to minimize the reflection coefficient. In the rest of the paper, we choose the taper to have a length of $4\;\mathrm{\upmu m}$ set at an angle of $40\degree$ and an end width of $200\;\mathrm{nm}$. Reflectivities are given at $\lambda_{\mathrm{SIP}}$.}
    \label{fig:Termination}
\end{figure}

At each side of the 3PD waveguide, we use the top waveguide for input and output ports. The other two waveguide extents are designed to minimize the reflection coefficient $\Gamma$ from a waveguide boundary in a relatively small use of space. In other words, we are looking to achieve $\Gamma = 0$ for two waveguides on each side in Fig.~\ref{fig:Geometry-3PD}. The nominal width at the waveguide/termination boundary is tapered to a $200\;\mathrm{nm}$ width at the end of the taper. In Fig.~\ref{fig:Termination}, we present a collection of simulation results across a subset of the termination taper design space. By changing the taper angle, length, and minimum width, we are able to reduce the reflected signal from each termination near the SIP frequency. Depending on the application and purpose of the measurement, either a large reflection coefficient (as was done in previous work with the 3PD) or minimal reflection coefficient may be beneficial.

\begin{figure}
    \centering
    \begin{subfigure}{0.42\textwidth}
        \includegraphics[width=\linewidth]{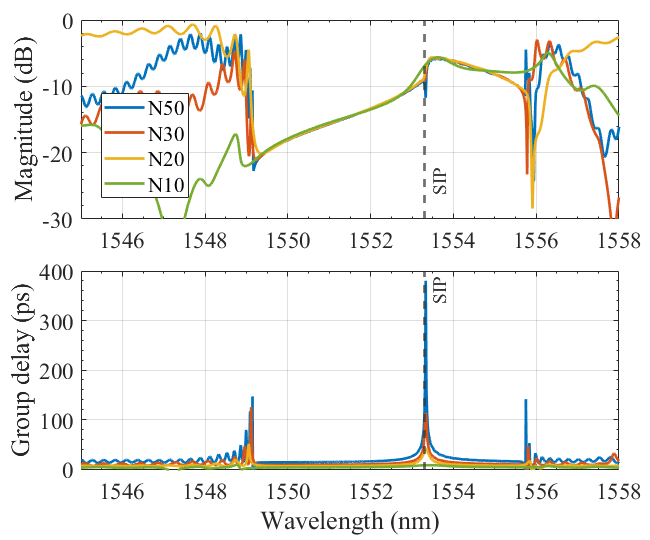}
        \caption{}
    \end{subfigure}
    \\
    \begin{subfigure}{0.42\textwidth}
        \includegraphics[width=\linewidth]{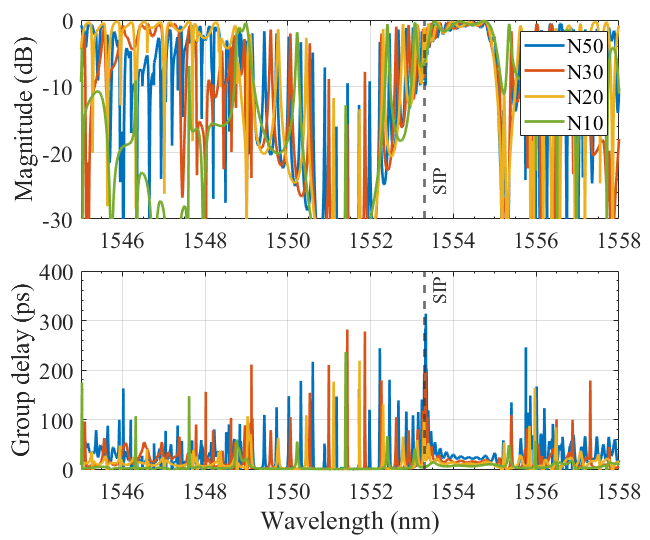}
        \caption{}
    \end{subfigure}
    \\
    \caption{Finite-length waveguide transfer function and group delay for the 3PD waveguide for various numbers of unit cells $N$. A reflection coefficient of $\Gamma = 0$ is shown in (a) and $\Gamma = 1$ is shown in (b).  By reducing the reflection coefficient at the waveguide boundaries we can more clearly distinguish the SIP associated resonance and group delay.}
    \label{fig:Transfer-Function-3PD}
\end{figure}

With that in mind, we particularly consider the dispersion and spectral response for the 3PD waveguide. The SIP appears to form inside of a pseudo bandgap (the region between the two RBEs). This is not a true bandgap--such as would be seen next to a traditional RBE or DBE--due to a  mode still propagating at all wavelengths (the same mode that has the SIP). This propagating mode corresponds to the center black line in Fig.~\ref{fig:Dispersion} for imaginary-valued wavenumbers. In finite-length models, we frequently observe a sharp transmission valley between the two adjacent RBEs around the SIP. We clearly observe this in the spectral response, given in Fig.~\ref{fig:Transfer-Function-3PD}(a), for small $N$ numbers. Importantly, as the finite-length device becomes longer, we start observing a resonance inside the pseudo bandgap with a corresponding anomalous scaling of group delay at that resonance. This resonance close to the SIP frequency in what is otherwise a transmission valley appears to be an indication of SIP behavior in the finite-length waveguides if measured alongside enhanced group delays.

We contrast the spectral response with a small reflection coefficient at the boundaries to the same response when using a large reflection coefficient of $\Gamma = 1$ in Fig.~\ref{fig:Transfer-Function-3PD}(b). The high reflectivity at the boundaries lead to a significant increase in the number of Fabry-P\'erot cavity resonances across the entire shown spectral range for all numbers of unit cells. We still see increases in group delay around the SIP and nearby RBEs, however it is much more difficult to differentiate SIP resonances from the multitude of other resonances within a nanometer spectral range or less of the SIP wavelength. The resolution of measurement instruments also comes into play over such a small wavelength range. The measurement device must have sufficient precision to identify two adjacent resonances in the wavelength range of interest. A notable improvement for the $\Gamma = 1$ case is the transfer function magnitude around the SIP frequency. With lossy terminations of $\Gamma \approx 0$, the magnitude of the SIP resonance is between $\shortminus 5$~to~$\shortminus 8\;\mathrm{dB}$ whereas for highly reflective terminations the magnitude is around unity at $0\;\mathrm{dB}$. In a linear scale, this equates to $\approx 50-60\%$ reduction in field magnitude. For some applications, this significant reduction in power is detrimental. However, we are interested in observing photonic waveguides operating around the SIP, making a lossy termination the clear choice for better defined spectral characteristics associated with the SIP and FMR. With $\Gamma \approx 0$, the single mode in the pseudo bandgap sees a lossy 3PD waveguide termination. However, when approaching the SIP, the degenerate Bloch mode tends to have a degree of mismatch that is responsible for the narrow resonant peak we still observe in the presence of lossy terminations.

A significant difficulty in predicting finite-length waveguide behavior for long devices is numerical instability in models. For the 3PD waveguide, the finite-length transfer function is calculated based on a cascaded unit cell transfer matrix model. Solving the finite-length problem requires inverting a $12\mathrm{x}12$ matrix with values ranging over fifteen orders of magnitude as $N$ increases. Over various spectral regions, the reciprocal condition number (a measure of the condition of a matrix under some small perturbation \cite{golubMatrixComputations1983}), degrades to near zero. This degradation negatively impacts the ability to accurately represent the spectral response and is increasingly noticeable for large-$N$ devices. Thus, we fabricated and measured some devices yet do not have as associated simulation comparison for the spectral response and group delay.

\section{Experimental Design}
\label{ch:Setup}
The devices reported here were fabricated in a commercial $300\;\mathrm{mm}$ CMOS foundry, the American Institute for Manufacturing Integrated Photonics~(AIM Photonics), through their multi-project wafer~(MPW) offering \cite{fahrenkopfAIMmpw2019}. We included the 3PD design on the fabricated photonic chips with lengths $N = 10, 20, 40, 60, 150$. All $N$ numbers include all termination geometries for all devices. The choice of $N$ is important in characterizing waveguide behavior. If the total waveguide length is too small (equivalently $N$ is too small), it is unlikely devices will operate around the FMR or have observable properties of the FMR. As the device length increases, challenges in modeling and fabrication disorder arise. Initial experimental investigations suggest fabrication disorder, i.e. non-uniformity between unit cells, does indeed negatively impact waveguide performance with larger $N$. Further, our investigations suggest a good accuracy of numerical models (with the exception of compounding numerical error in some spectral regions for large $N$). Thus, in this ongoing experimental study, we pushed device lengths to the millimeter range while still including sub-millimeter lengths. A key change, as mentioned in the proceeding section, to enable the measurement of longer devices is reducing the reflection coefficient at the waveguide boundaries to $\Gamma \approx 0$. Without this updated termination tapering, our characterization of measured data would be hindered.

Previous studies suggest waveguides operating near band-edge degeneracies display observable properties of the degeneracy at much smaller $N$. For instance, the authors in Ref.~\cite{burrExperimentalVerificationDegenerate2016} fabricate a DBE resonator with $N = 35$ and $d = 390\;\mathrm{nm}$ and excite it around $\lambda = 1541\;\mathrm{nm}$. The total device length is smaller than $10 \lambda$, and this scale of multiple-$\lambda$'s is common in related works. Indeed, as shown in Fig.~\ref{fig:Transfer-Function-3PD} for $N$ as low as $10$, the band edges of the next closest RBEs are visible. Herein also lies a tradeoff between design geometries and length. Comparing other waveguide geometries with a considerably smaller unit cell size than the 3PD (such as in Ref.~\cite{burrExperimentalVerificationDegenerate2016}), it is reasonably expected to see SIP-associated behavior in an overall smaller footprint.

\begin{figure}[b]
    \centering
    \frame{\includegraphics[width=0.48\textwidth]{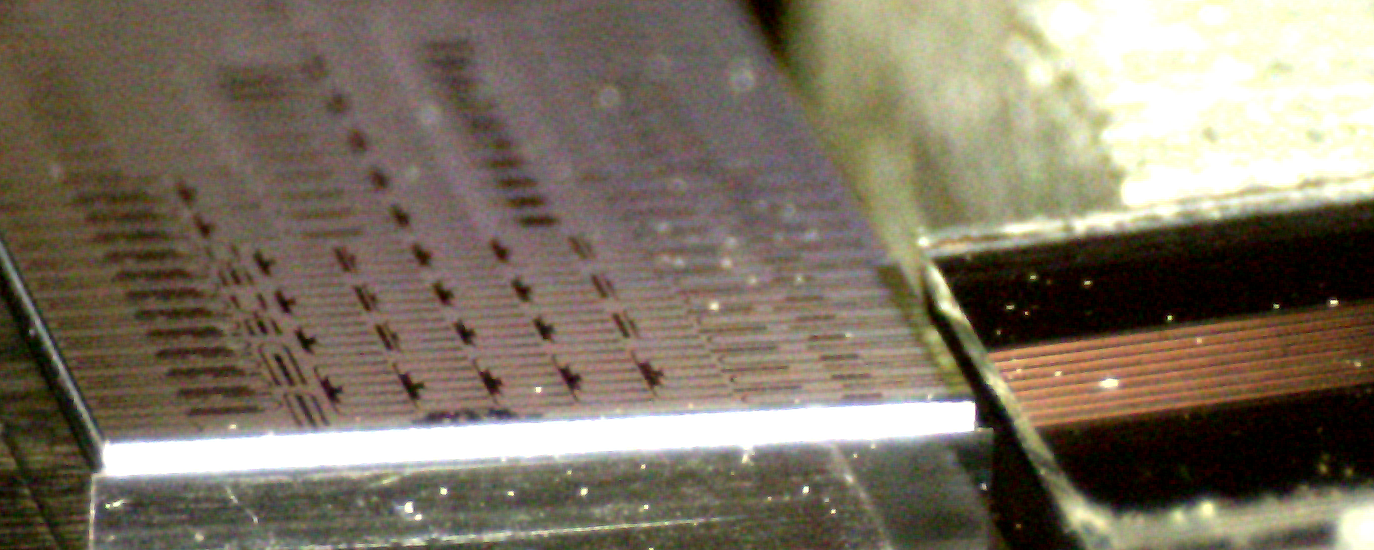}}
    \caption{Example measurement setup with a photonic chip on the left and V-groove fiber array on the right. Light is edge-coupled into the chip from the fiber array connecting to the measurement system.}
    \label{fig:Edge-Coupled-Measurement}
\end{figure}

We measured finite-length devices on six individual photonic chips using a Luna Optical Vector Analyzer~(OVA) 5100. This system acts as both the source and detector and works based on time-of-flight data from an internal broadband source. The system has a resolution of $\pm 1.5\;\mathrm{pm}$ and the presented data is measured at this maximum reported resolution \cite{LunaOVA}. The wavelength accuracy of this system is in the $1-2\;\mathrm{nm}$ range. A key feature of this system is its ability to provide information on the group delay. It is important to note the group delay from the OVA system is for the entire measurement path--including connecting fiber cables and on-chip paths. The basic operating principle of the OVA is measuring the transmission return power of a broadband and polarized laser over time. Based on this time signal, the system calculates the transmission and delay based on the time-of-flight response along with a Jones polarization matrix calculation. As this delay includes the total distance along the measurement path, it is necessary to subtract the `baseline' delay to only see the response from the device itself. The photonic chip includes half-circle cutback paths (diameter of $127\;\mathrm{\upmu m}$) without any devices to characterize this baseline level of $\approx 3.123\;\mathrm{ns}$ delay. Figure~\ref{fig:Edge-Coupled-Measurement} shows an example measurement of a photonic chip setup where light is edge-coupled to the chip using a V-groove fiber array. As is seen in this figure, the input and output paths are along the same side of the chip. Thus, a waveguide return path is necessary next to the device-containing path. These return paths have a width of $w = 480\;\mathrm{nm}$, and using the modal index calculated from simulations of $n_w = 2.378$, the delay per length in this straight waveguide is $\approx 7.9\;\mathrm{fs/\upmu m}$. We include a $20\;\mathrm{\upmu m}$ segment to gradually change the waveguide width from $450\;\mathrm{nm}$ to $480\;\mathrm{nm}$ as not to induce extra reflections from the width change. Thus, the presented group delay data has both the baseline delay and the delay calculated from the waveguide return path subtracted out as to only see the delay from the devices themselves. Specifically referring to Fig.~\ref{fig:Geometry-3PD}, we subtract out the delay for all paths outside the $z=0$ to $z=Nd + d_{ex}$ range.

Each chip is denoted by an identifying number from the foundry (i.e., C15, C20, etc.). On each photonic chip, we have included an additional device to aid in characterizing the frequency response of the DBRs used in the 3PD waveguide unit cell. The DBR used in the 3PD waveguide has $M = 12 + 8$ unit cells with periods of $365\;\mathrm{nm}$ (not to be confused with the unit cell of the 3PD waveguide with period of $27.0\;\mathrm{\upmu m}$) and an outer width of $450\;\mathrm{nm}$. Twelve of these unit cells have an inner width of $350\;\mathrm{nm}$ while the four on each end are apodized to reduce scattering losses. We optimized the period of the DBR to provide maximum reflection at $\lambda_{\mathrm{SIP}}$ over the length $L_{DBR}$. This twenty period device used in the 3PD waveguide unit cell only results in a transmission decrease of $\approx 1.5\;\mathrm{dB}$. To cleanly measure the spectral features of this DBR, we cascaded the $365\;\mathrm{nm}$ period unit cell $M = 3,000$ times in the fabricated devices. This long DBR provides high reflectivity with sharp transmission drop-offs near the upper and lower cutoff frequencies. Instead of using the OVA system to measure this characterization DBR structure, we instead use a higher power superluminescent diode~(SLD) with spectra centered around $\lambda_{\mathrm{SIP}}$ paired with an optical spectrum analyzer~(OSA). The group delay is not needed for the long DBRs so this more versatile SLD/OSA setup is suitable for our analysis needs.

For easier visual comparison of 3PD measured spectral response data, we calculate the cross-correlation~(XC) of the measured data with respect to a reference chip and normalize measured data to $0\;\mathrm{dB}$ across the $\approx 1530-1580\;\mathrm{nm}$ range. The XC is generally used in audio signal processing and pattern recognition \cite{rabinerTheoryApplicationDigital1975,bracewellFourierTransformIts1965}. From the XC, we determine the nanometer wavelength offset, defined as the XC value where the maximum magnitude occurs within a subset of the overlapping spectral range. The presented measured data in Fig.~\ref{fig:nmShift} is thus spectrally shifted to align with the chip C15 reference measured results.

\section{Experimental Results}
\label{ch:Results}

\begin{figure}
    \centering
    \includegraphics[width=0.42\textwidth]{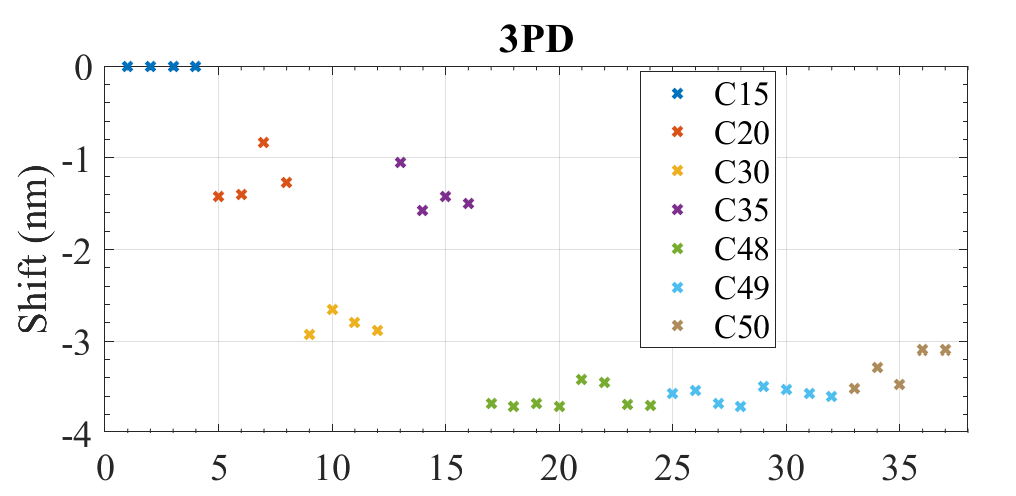}
    \caption{Spectral shift of the 37 measured 3PD waveguide devices given in nanometers, calculated based on the cross correlation with chip C15. Each `x' represents an individual measurement of a finite-length structure.}
    \label{fig:nmShift}
\end{figure}

We measured the spectral response data for six different photonic chips (C15, C20, C30, C35, C48, and C49) using the OVA system and we used the SLD/OSA system to measure the response for all other chips. Across the different chips and for varying $N$, the overall spectral responses agree reasonably well over a wide spectral range. Comparing the responses chip-to-chip, we observe a spectral wavelength shift (calculated from the XC) around $3-4\;\mathrm{nm}$ between the measured chips. In Fig.~\ref{fig:nmShift}, we plot the nanometer offset for the 3PD waveguide when using chip C15 as the reference chip (i.e., zero shift from the autocorrelation). Each `x' represents an individual measurement. Here, chips C48, C49, and C50 have similar spectral offsets, suggesting that the spectral responses of these chips are more closely aligned. This few nanometer variation is reasonable and expected across an entire wafer-sized area. While we use chip C15 as the reference, we could have equivalently used any of the other chips instead. Using a different reference would only change the values of the shift-per-chip rather than their distribution.

\begin{figure}
    \centering
    \includegraphics[width=0.42\textwidth]{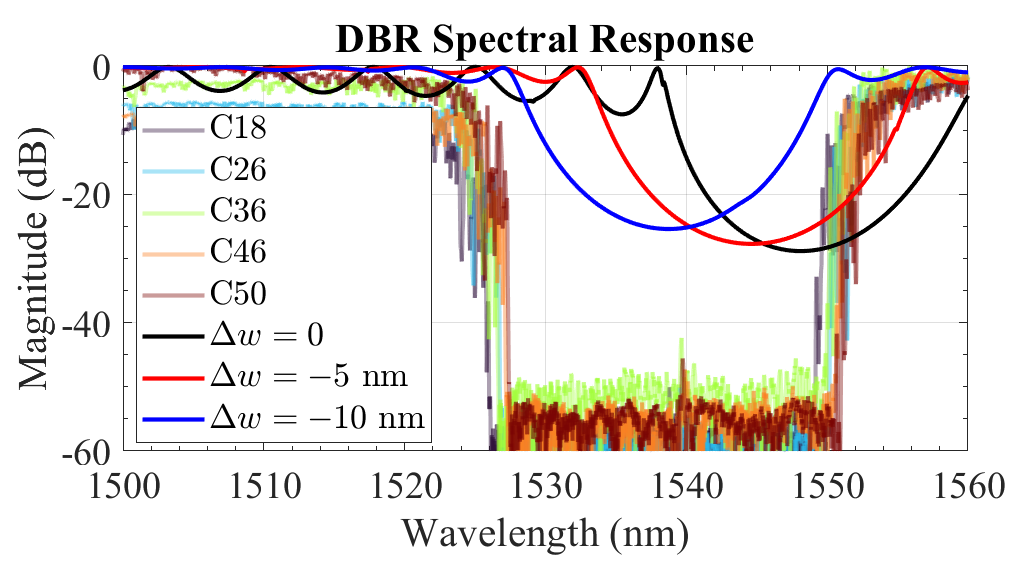}
    \caption{Measured spectral response of the $M = 3,000$ DBR for different chips (plotted in various colors), and simulated response of the $M = 100$ DBR under no waveguide width perturbation (plotted in black), under a $\shortminus 5\;\mathrm{nm}$ width perturbation (plotted in red), and under a $\shortminus 10\;\mathrm{nm}$ width perturbation (plotted in blue). The period of an individual DBR unit cell is $365\;\mathrm{nm}$ and the nominal width is $450\;\mathrm{nm}$.}
    \label{fig:SpectralResponseDBR}
\end{figure}

We continue by highlighting the measured spectral response for the extended, long DBRs in Fig.~\ref{fig:SpectralResponseDBR}. This data is the raw data and has no wavelength shift applied. The XC wavelength shift is not plotted for this data, yet we calculate the overall shift to be in the $\approx 1.5\;\mathrm{nm}$ range, half of that for the 3PD waveguide finite-length device. The measured data shows a transmission valley from $\approx 1525-1552\;\mathrm{nm}$, just shorter than the $\lambda_{\mathrm{SIP}} = 1553\;\mathrm{nm}$ mark. To better understand this offset from the expected DBR spectrum, we calculate the transmission response from full-wave simulations of $M = 100$ unit cells (rather than the $M = 3,000$ included on the photonic chip). This smaller simulation geometry still gives useful information while not requiring overly intense computational resources. The black curve in Fig.~\ref{fig:SpectralResponseDBR} is the nominal DBR (i.e., $365\;\mathrm{nm}$ period, outer/inner widths of $450/350\;\mathrm{nm}$). As we have done in previous studies to better understand the effects of modal index changes $\Delta n_w$, we let a change in waveguide width $\Delta w$ represent the variety of physical difference between simulation and fabrication, including material effective permittivity changes or waveguide geometric cross-section changes. The red and blue curves in Fig.~\ref{fig:SpectralResponseDBR} represent additional simulations where we have decreased \textit{both} the outer and inner waveguide width by the same amount of $5\;\mathrm{nm}$ and $10\;\mathrm{nm}$ shown with red and blue curves, respectively. The DBR unit cell period length and duty cycle are unchanged.

By decreasing the waveguide width in full-wave simulations, we reduce the effective index of the waveguide mode and blue-shift the DBR spectral response to smaller wavelengths. The black curve for the nominal response is centered around $1550\;\mathrm{nm}$, as expected. We again note that the smaller twenty period DBR used in the 3PD waveguide unit cell still has its maximum reflectivity around $\lambda_{\mathrm{SIP}}$. When we decrease the long $M = 100$ DBR waveguide width to $\shortminus 10\;\mathrm{nm}$ in simulations, the expected spectral response agrees much better with the measured spectral response. This difference in the nominal simulation to measured data suggests that the fabricated devices may have a smaller modal index than initially predicted.

\begin{figure}
    \centering
    \includegraphics[width=0.42\textwidth]{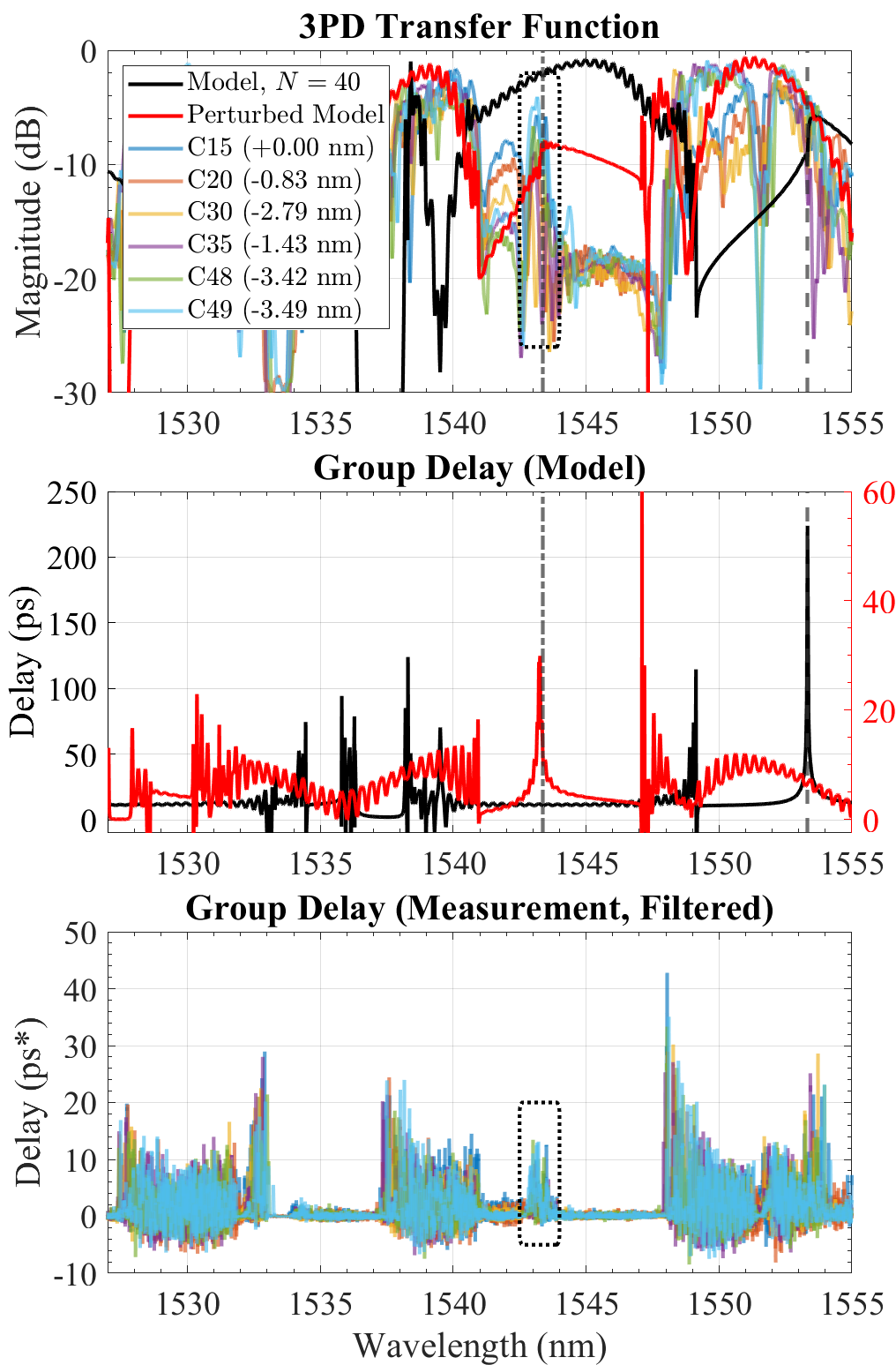}
    \caption{Transfer function and group delay for the 3PD waveguide with $N=40$ unit cells. Black lines represent the expected, or modeled, response while measurements for six individual photonic chips are shown in various other colors. The measured data has been spectrally aligned with respect to chip C15, calculated from the XC, displayed in the legend of the top plot. We filter the group delay for measured data from a multiplication of the delay with the magnitude of the transfer function (in a linear scale), given as $\tau_{g,f} = \tau_g \times |T_f|$. The vertical gray dashed line in the top two plots mark the wavelength corresponding to the maximum delay, nearly identical to the SIP wavelength. The rectangle with black dashed lines in the top and bottom plots highlight the region of interest for measured data where we see enhanced group delay responses. The red curve, with its peak delay in the extents of the rectangle marked by a vertical gray dashed-dot line, represents the transfer function and group delay of the model under some waveguide width perturbation. We spectrally shift the red curve by $+2.50\;\mathrm{nm}$ (both transfer function and group delay).}
    \label{fig:SpectralResponse3PD-N40}
\end{figure}

Now looking at the spectra for the finite-length waveguides, we display the modeled and measured spectral responses for the 3PD waveguide with $N = 40$ unit cells in Fig.~\ref{fig:SpectralResponse3PD-N40}. We identify the chips by the same numbers as before and display their corresponding XC wavelength offset in the figure legend. Using Eq.~\ref{eq:GroupDelay}, we calculate the group delay and plot the result in the middle figure and we plot the delay from measurements of the same waveguide on various chips in the lower figure. We show the group delay for measurements as the filtered delay and we also spectrally shift the data the same way as the spectral response. We calculate the filtered delay from a simple multiplication of the unfiltered delay with the magnitude of the transmission response as $\tau_{g,f}=\tau_g \times |T_f|$. This filtering is useful in differentiating actual delay peaks from `phantom' delay peaks resultant from low signal as detected by the OVA. Reduction in transmission causes errant spiking of the unfiltered delay, which at times imparts numerical error. The red curve in this figure represents the transfer function of a ``perturbed'' model and we discuss this model shortly.

\begin{figure}
    \centering
    \begin{subfigure}{0.42\textwidth}
        \includegraphics[width=\linewidth]{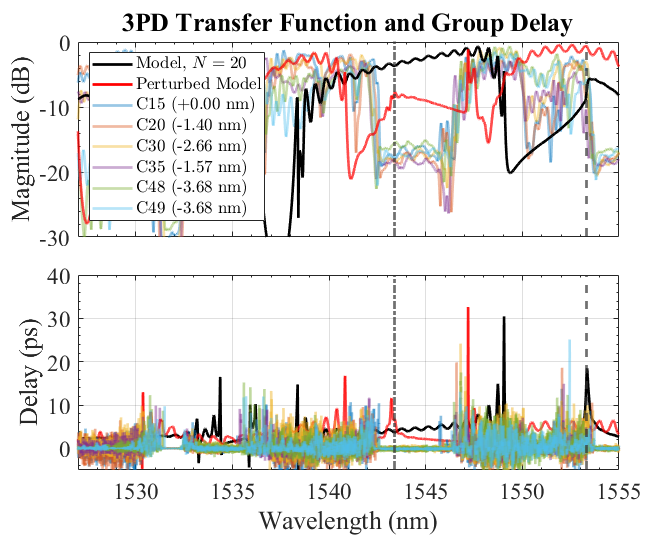}
        \caption{}
    \end{subfigure}
    \\
    \begin{subfigure}{0.42\textwidth}
        \includegraphics[width=\linewidth]{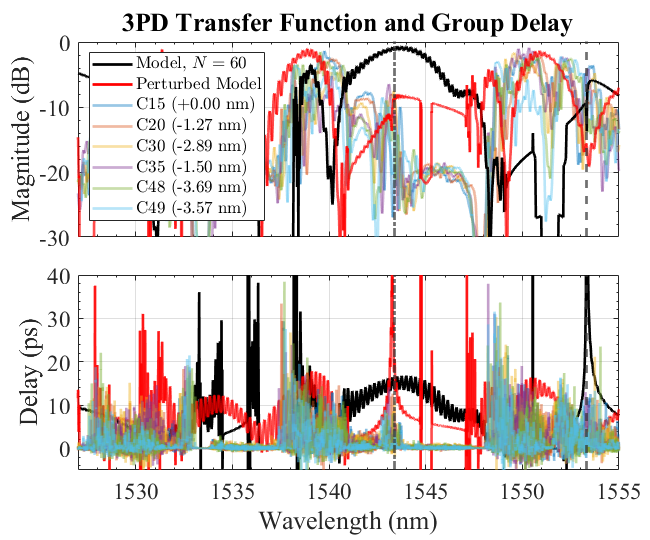}
        \caption{}
    \end{subfigure}
    \caption{Similar and more compact plots to those in Fig.~\ref{fig:SpectralResponse3PD-N40}, shown here for (a) $N = 20$ and (b) $N = 60$. The red plots are perturbed by the same amount in every plot. The measurement delays are filtered while the modeled delays are not. The sharp transmission valleys in (b) around $1552\;\mathrm{nm}$ for the black curve and around $1545\;\mathrm{nm}$ for the red curve are due to numerical instability for large $N$ devices.}
    \label{fig:SpectralResponse3PD-Other}
\end{figure}

We observe large group delays over a few wavelength ranges in Fig.~\ref{fig:SpectralResponse3PD-N40}. The group delay from the model in Fig.~\ref{fig:Transfer-Function-3PD}(a) shows a large spike of delay at the SIP and at the two adjacent band edges, paralleled for the black curve here in Fig.~\ref{fig:SpectralResponse3PD-N40}. When considering the filtered delay for \textit{measured} data (bottom plot), there appear to be delay increases around wavelengths of $1541\;\mathrm{nm}$, $1543\;\mathrm{nm}$, and $1549\;\mathrm{nm}$, corresponding to resonances close to the two RBEs. When the measured spectral response displays a resonant peak, such as around $\lambda = 1543-1544\;\mathrm{nm}$ (highlighted by a dashed-line rectangle in the top plot), yet has a reduced transmission at such resonances, the filtered delay can be somewhat misleading. In this particular case, the spectral resonances are adjacent to spectral valleys so the filtered delay on-resonance is visually well isolated from the nearby transmission valleys. When observing the group delay in the region marked by the black rectangle, the filtered delay shows notable spikes (around $15-18\;\mathrm{ps}$) yet those spikes correspond to magnitudes less than $\shortminus 5\;\mathrm{dB}$. The raw, unfiltered delay, is thus closer to the $35-45\;\mathrm{ps}$ range. We show compacted data plots for $N = 20$ and $N = 60$ in Fig.~\ref{fig:SpectralResponse3PD-Other} as well.

\begin{figure}
    \centering
        \includegraphics[width=0.42\textwidth]{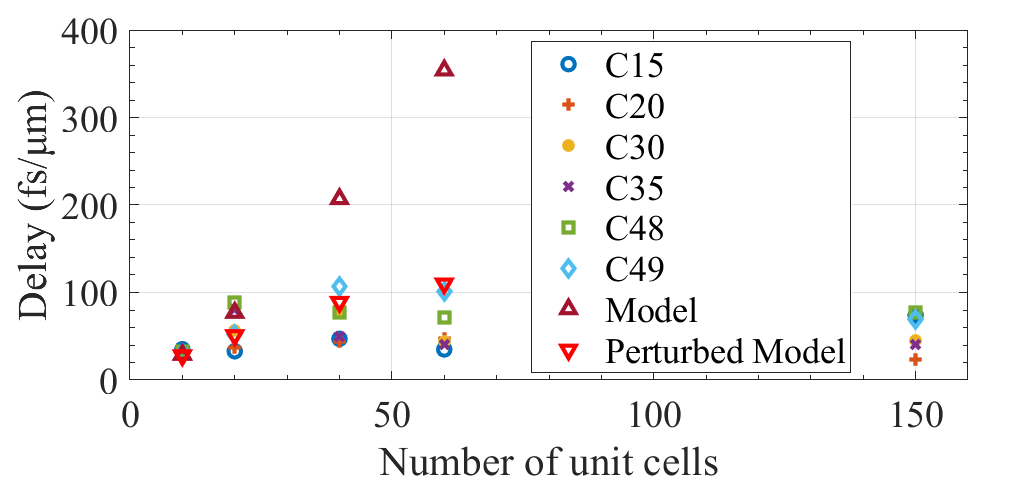}
    \caption{Group delay per unit length ($\mathrm{fs/\upmu m}$) for the 3PD waveguide. The expected delays (denoted by model dark red triangle) are given alongside the measured delays (denoted by chip number) from the OVA system where both are chosen based on the peak delay at resonances around the SIP frequency. The measured delays take into consideration the background round trip delay from connected measurement equipment (nanosecond range delays) and on-chip connecting paths (picosecond range delays). We also plot the perturbed model delays in bright red for up to 60 unit cells.}
    \label{fig:Delay-3PD}
\end{figure}

Plotting and comparing group delays are important, yet generally as devices become longer, their corresponding group delays increase. It is more useful, therefore, to compare the group delay per unit length. This allows for more straightforward comparisons of different devices, designs, and waveguides, and further can give insight into the effects of fabrication disorder in these waveguides. In Fig.~\ref{fig:Delay-3PD} we plot the normalized delays for the 3PD waveguide. These delays per length are for the devices only, meaning that we accounted for the cutback return paths and other outside measurement equipment delays and subtracted the non-device delays from the measured group delay.

As mentioned previously, compounding numerical instability thwarts attempts for accurate estimates of group delay in analytic models for long waveguides. Thus, for $N = 150$ in Fig.~\ref{fig:Delay-3PD}, we only present the delay from measured data. As each chip has variable fabrication disorder, and further the model and measurement devices may be slightly mismatched, it is interesting to observe how many, and for which $N$, chips follow the expected trend for delay per unit length. In particular, the modeled and measured delays per length are comparable for $N = 10$ (near the background $\approx 7.9\;\mathrm{fs/\upmu m}$ value) and more importantly for $N = 20$. Naturally, there is some variation among chips, with a few instances of the delay from measured data being larger than that for modeled calculations. Observation of a longer delay than expected is not unreasonable (such as for the delay of C48 for $N = 20$), as previous studies working with representative fabrication disorder have shown. It is important to note that the \textit{total} delay for $N = 60$ is greater than for $N = 40$, even though the delay per length is comparable. As the devices become longer, we continue to observe larger group delays yet not the expected increase in the delay per length. We discuss these differences in the next section. We further note that the decrease in delay per length from measured data is unlikely due to losses in the waveguide. We continue to observe well-defined spectral features for large $N$ and have taken steps to reduce losses in the unit cell geometry.

\section{Discussion}
\label{ch:Discussion}

The measured spectral response for the presented 3PD waveguide shows notable resemblance to the expected modeled response, albeit with about a $\shortminus 10\;\mathrm{nm}$ shift in wavelength. In particular, and as predicted from the modeled finite-length results in Fig.~\ref{fig:Transfer-Function-3PD}, the SIP resonance and corresponding increase in group delay are expected to occur inside a pseudo bandgap defined by the two nearest RBEs, one above and one below the SIP (where a single mode propagates). The spectral response resonances in measured data (Figs.~\ref{fig:SpectralResponse3PD-N40}~and~\ref{fig:SpectralResponse3PD-Other}) around $1543-1544\;\mathrm{nm}$ do indeed fall inside a pseudo bandgap between $1541-1548\;\mathrm{nm}$. In fact, even around the edges of the bandgap, the modeled and measured responses are notably similar. With paralleled sharp transmission valleys, the analytic model well approximates measured data. These similar spectral responses accordingly translate to the modeled and measured group delays. The measured group delay is enhanced around the spectral response resonances seen around $1543-1544\;\mathrm{nm}$, highlighted by the dashed-line rectangle in the top and bottom plots of Fig.~\ref{fig:SpectralResponse3PD-N40}. Further, there appears to be spiking of the delay at each edge of the pseudo bandgap around $1541\;\mathrm{nm}$ and $1548\;\mathrm{nm}$. We see analogous observations for the lengths of $N=20$ and $N=60$ in Fig.~\ref{fig:SpectralResponse3PD-Other}. The measured group delay also follows a similar pattern to the modeled delay with an increase inside the pseudo bandgap along with increases at the surrounding band edges.

\subsection{Reverse-Modeling}
To better understand the discrepancies between the model and fabricated devices, we refine model variations to explore systematic changes between design and fabricated devices across different MPW chips (as opposed to investigating fabrication disorder between unit cells of a single devices). Towards that end, we consider how perturbations in the waveguide width shift the modeled spectral response (both transmission and group delay). As was done in prior work (see Ref.~\cite{furmanImpactWaveguideImperfections2023} for instance), a change in the waveguide width corresponds to a change in the straight waveguide and coupled waveguide modal indices. For a width change on the order of a few nanometers, $n_w$ changes by less than a few parts in a hundred. Changes in $n_w$ and in the even/odd mode modal indices $n_e$, $n_o$ are understood through a change in the waveguide width (and corresponding change in the gap size between waveguides), yet the physical mechanism for the change could be resultant from different material relative permittivities, different waveguide cross-section geometries, or more.

From our analysis of the XC wavelength shift of the response of 3PD devices and of the the long DBR structure, we notice that the long DBRs have a more consistent spectrum chip to chip compared to the more complicated finite-length devices. This XC wavelength shift difference suggests that disorder applied to the overall 3PD waveguide unit cell may not accurately capture the different DBR perturbation. We therefore expand the perturbation analysis to consider overall unit cell width perturbations individually from DBR width perturbations. To confine the parameter space, we consider changes of the widths by multiples of two nanometers between the nominal width and $\Delta w = \shortminus 10\;\mathrm{nm}$, resulting in $36$ combinations of overall waveguide width and DBR width perturbations. As we did for the simulation DBR in Fig.~\ref{fig:SpectralResponseDBR}, we decrease both the outer and inner waveguide widths while keeping the DBR period and duty cycle the same.

To better characterize the perturbation analysis beyond simple comparison with the measured transmission spectra, we also consider the perturbed dispersion diagram associated with the unit cell geometry. This dispersion diagram is calculated the same way as done in Fig.~\ref{fig:Dispersion} for the nominal 3PD waveguide except we update both the waveguide width and the DBR widths. We use DBRs in the perturbed 3PD waveguide unit cell based on full-wave simulation methods just as done in the nominal case. We also re-introduce the coalescence parameter $C$ from Refs.~\cite{furmanFrozenModeRegime2023,herrero-parareda_frozen_2022}. This parameter is a measure of how close three modes are to becoming degenerate (i.e., coalescing) at a given frequency, with a smaller value representing a better coalescence. At a perfect SIP $C = 0$, and $C$ is always between zero and one. When analyzing the perturbation space, we use the coalescence parameter as a tool to guide applied perturbations in tandem with the finite-length waveguide spectral response.

\begin{figure}
    \centering
    \includegraphics[width=0.48\textwidth]{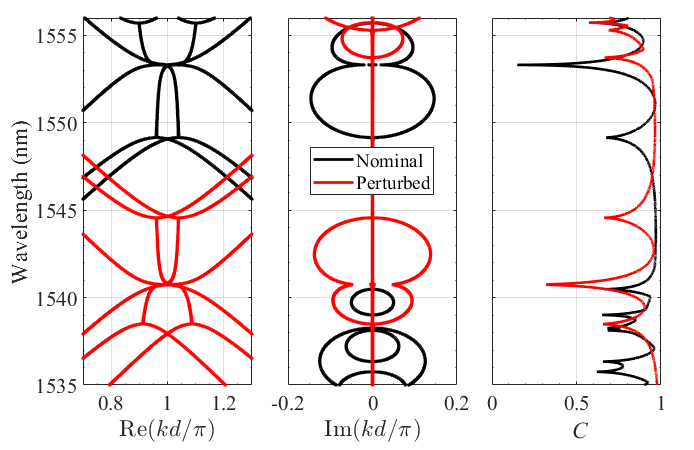}
    \caption{Dispersion diagram and coalescence parameter of the modes in the 3PD waveguide. The black curves are the nominal dispersion (identical to Fig.~\ref{fig:Dispersion}) and the red curves are the perturbed dispersion. The coalescence parameter is a minimum for the nominal response at $\lambda_{\mathrm{SIP}} = 1553.3\;\mathrm{nm}$ and is a minimum for the perturbed response at $1540.9\;\mathrm{nm}$.}
    \label{fig:Dispersion3PD-Perturbed}
\end{figure}

In Fig.~\ref{fig:Dispersion3PD-Perturbed}, we plot the dispersion diagram of the 3PD waveguide unit cell for the nominal case in black (same curve as the one in Fig.~\ref{fig:Dispersion}(a)) and for the perturbed case in red. We also plot the coalescence parameter for both cases. Here, we perturb the waveguide width by $\shortminus 8\;\mathrm{nm}$ (corresponding to a reduction in the modal index of $\Delta n_w = \shortminus 0.022$) and we perturb the DBR widths by $\shortminus 4\;\mathrm{nm}$ (both outer and inner widths, corresponding to a transmission valley centered at $\approx 1551.7\;\mathrm{nm}$). This combination of waveguide and DBR perturbation still results in a small coalescence parameter indicating vicinity to an SIP. As expected, we see the coalescence parameter of the nominal response experience a minimum at $\lambda_{\mathrm{SIP}} = 1553.3\;\mathrm{nm}$. The perturbed response shows a coalescence parameter minimum of a similar (but not equal) magnitude at $1540.9\;\mathrm{nm}$. This $12.4\;\mathrm{nm}$ wavelength difference corresponds to a separation of $\approx 1.55\;\mathrm{THz}$. While no longer a perfectly coalescing SIP, the perturbed dispersion response clearly shows SIP-like behavior at these smaller wavelengths. The perturbed dispersion diagram additionally still shows spectrally adjacent RBEs which correspond well to the transmission bandgap observed in the analytic and measurement transfer functions.

We plot the finite-length spectral response under the same perturbation as the red curves in Figs.~\ref{fig:SpectralResponse3PD-N40}~and~\ref{fig:SpectralResponse3PD-Other}. We also spectrally shift the red curves by $+2.5\;\mathrm{nm}$ to better align with measured data. In other words, we observe the minimum coalescence at $1540.9\;\mathrm{nm}$ now at $1543.4\;\mathrm{nm}$. Considering Fig.~\ref{fig:SpectralResponse3PD-N40}, we observe a significantly better agreement of the perturbed model with respect to the measured data. This agreement is mirrored for both $N = 20$ and $N = 60$ in Fig.~\ref{fig:SpectralResponse3PD-Other} where the plotted red curves have the same unit cell perturbations. Further, and as stated in the previous section, based on calculations of the XC wavelength shift we see a smaller variation of the XC wavelength shift for the long DBRs compared to the full devices. The smaller DBR width perturbation compared the overall 3PD waveguide width perturbation well corresponds to the observed shifts. The XC wavelength shift between the collection of long DBRs is half that for the full waveguide devices ($1.5\;\mathrm{nm}$ shift variation compared to $3\;\mathrm{nm}$) and the DBR width perturbation is half that for the overall width perturbation ($\shortminus 4\;\mathrm{nm}$ compared to $\shortminus 8\;\mathrm{nm}$). The chosen width perturbations for both the DBR and rest of the waveguide widths are also within reasonable ranges when considering the measurement and simulation data from the long DBR in Fig.~\ref{fig:SpectralResponseDBR}. A change of $\Delta w = \shortminus 10\;\mathrm{nm}$ results in a wavelength response shift of $\approx 12\;\mathrm{nm}$, comparable to the differences in dispersion and finite-length spectral responses between modeled and measured data. However, since we see a wavelength shift less than $12\;\mathrm{nm}$ between the nominal model and measured data, it follows that a $\Delta w = \shortminus 10\;\mathrm{nm}$ is also on the upper end of the perturbation space.

\begin{figure}
    \centering
    \includegraphics[width=0.48\textwidth]{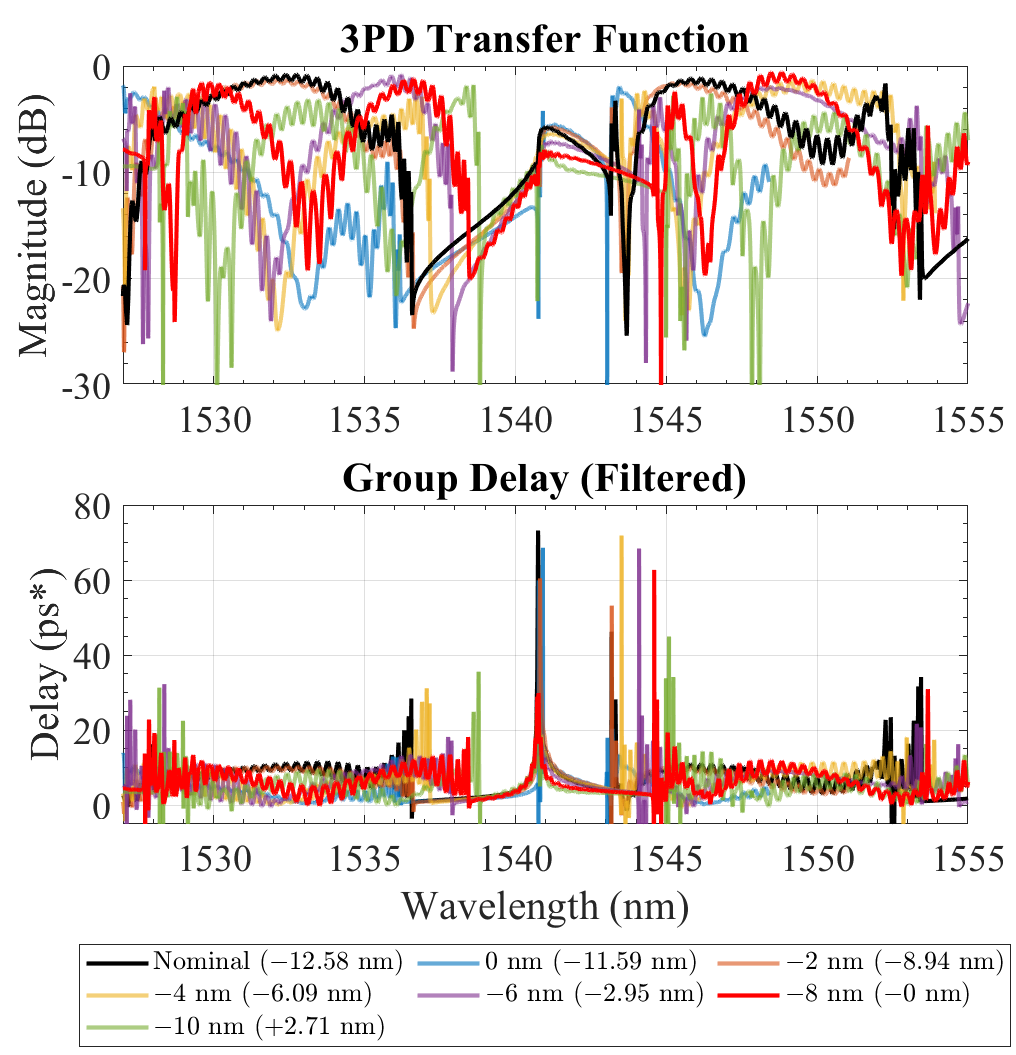}
    \caption{Transfer function and group delay (filtered) of the 3PD waveguide for a varying waveguide width (numbers outside the parentheses). In all cases the DBR width perturbation is fixed to $\shortminus 4\;\mathrm{nm}$ (both outer and inner width). Spectral shifts are applied (value in parentheses) to align the partial SIP occurrences with that of the red curve. The nominal response is plotted in black, the same corrected curve as in Fig.~\ref{fig:SpectralResponse3PD-N40} is identically plotted here in red, and each curve is spectrally shifted to align with the corrected red curve with the corresponding shift shown in the legend.}
    \label{fig:SpectralResponse3PD-Perturbed}
\end{figure}

To give a more complete description and elaborate on the perturbation search space, we fix the DBR width perturbation at $\shortminus 4\;\mathrm{nm}$ and apply different overall width perturbations to the finite length model in Fig.~\ref{fig:SpectralResponse3PD-Perturbed} for $N = 40$. The black curve is the nominal response and each $\Delta w = \shortminus 2\;\mathrm{nm}$ step is plotted in other colors (value in the legend outside of the parenthesis). Each of these curves are aligned with the red curve and are thus spectrally shifted (value in the legend inside the parenthesis). As we decrease the waveguide width, the transmission response gradually changes and shifts to shorter wavelengths. When $\Delta w = \shortminus 8\;\mathrm{nm}$ (the same as the red curves in previous transfer function figures), we see the best agreement between the perturbed model and measured data. The group delay under this width perturbation shows additional interesting information. As the width perturbation increases, the peak of group delay corresponding to the SIP resonance frequency decreases in magnitude. This makes sense when considering the model is deviating from the nominal response. Yet even with these perturbations, we continue to see significant group delay peaks and an overall similar spectral response. Unlike the red curves for the perturbed model in previous plots, here we apply no spectral shift to the perturbed model in Fig.~\ref{fig:SpectralResponse3PD-Perturbed}. The DBR transmission around its minimum at $1551.7\;\mathrm{nm}$ is $\shortminus 1.15\;\mathrm{dB}$, while at $1540.9\;\mathrm{nm}$ (the wavelength where $C$ is at a minimum) the transmission is only marginally different at $\shortminus 0.97\;\mathrm{dB}$.

Focusing again on the perturbed model in Figs.~\ref{fig:SpectralResponse3PD-N40}~and~\ref{fig:SpectralResponse3PD-Other} and when looking beyond the transfer function and at the group delay, we still see good agreement between the perturbed model and measurement. Both the model and measurement show increased delays around $1543-1544\;\mathrm{nm}$ in Figs.~\ref{fig:SpectralResponse3PD-N40}~and~\ref{fig:SpectralResponse3PD-Other}. Where we expect to see nearby RBE associated delay spikes in the model, we too see increases in delay in the measured data. By changing the modal indices by a small amount (via a waveguide/DBR width perturbation), the analytic model transfer function better matches the measured data. Not only do the transfer functions more closely match, the group delays are also more similar. The perturbed group delay response naturally shows a decrease in delay from the ideal peak around the nominal SIP wavelength, yet the peaks seen in this perturbed response are more in-line with the measured group delay. This slight tweaking of the analytic model reinforces the overall accuracy of numerical methods in calculating the expected transfer function and group delay; and furthermore because of this good accuracy, provides more evidence that these fabricated waveguides are operating near desired spectral degeneracies. While not a perfect SIP, the observed group delays are related to a flattening of the dispersion diagram in the wavelength range of interest and may indicate a kind of tilted SIP.

The increase in measured delay per length for longer device lengths is generally smaller than the increase in delay for shorter devices. For instance, in Fig.~\ref{fig:Delay-3PD}, on average the delay per length moderately increases for $N = 10$, $20$, and $40$. After $N = 40$, at $N = 60$ and $N = 150$, the delay per length does not continue to increase and is comparable to a device with $N = 10$ or $N = 20$. In Fig.~\ref{fig:Delay-3PD}, we also plot the delay per length for the perturbed model in red. As previously stated, the perturbed model is not a perfect SIP and thus we do not expect the same group delay scaling trend for large $N$ as we would for a perfect SIP \cite{nada_frozen_2021, furmanFrozenModeRegime2023,herrero-parareda_frozen_2022}. Comparing the nominal model to the perturbed model, we see up to hundreds of $\mathrm{fs/\upmu m}$ difference in the delay for larger $N$. However, the perturbed model delay per length is much more representative of the measured delays. Notably at $N=40$ and $N=60$, the perturbed model and the measured delay per length are quite similar. This trend further reinforces the accuracy of the perturbed model.

While the perturbed model and measured data are similar, measured data generally still has a smaller delay per length. Not only is the perturbed model slightly different than the physical device, fabrication disorder likely contributes to the reduction in delay. As the devices become longer and longer, modal index perturbations resulting from an imperfect fabrication process very likely contribute and compound to degradations in the measured spectral response and corresponding group delay increases. At small device lengths, the modeled and measured spectral responses are in good agreement and so too are the measured group delays. Only when devices become longer does fabrication disorder over the entire length of the waveguide negatively impact the measured spectral response and group delay. Further, as seen from the perturbation figures above, the group delay peaks are smaller than the nominal delay. It is then unsurprising that the measured delay per length also shows a decrease from the nominal values.

We reiterate the two primary distinctions of fabrication disorder: chip-to-chip disorder and unit cell-to-unit cell disorder. The differences between chips correspond to the $\approx 3\;\mathrm{nm}$ wavelength shifts seen in Fig.~\ref{fig:nmShift}. These are directly related to larger differences across an entire $300\;\mathrm{mm}$ silicon wafer. The intra-device disorder is seen at a smaller scale and contributes to differences in the phase, coupling coefficient, and more. Because of differences between the layout design and the fabrication process, the modal indices of the fabricated structures are slightly different than the nominal response. We attempt to capture these slight differences in our perturbed modeling approach. This approach does not directly consider the discussed types of disorder, yet does inform design differences and iterations.

Ultimately, based on the spectral response and group delay data from analytic models and measured chips, the presented device is likely to be operating in or around the frozen mode regime associated with the SIP. Not all chips for all numbers of unit cells, however, seen to be operating in this regime, yet a significant fraction of measured devices both have notable agreement of predicted and measured transfer functions with a corresponding increase of group delay at resonances around the SIP wavelength. Even without clear anomalous scaling of the group delay with waveguide length, the choice of boundary conditions for the 3PD has allowed for effective verification of analytic models suggesting operation near desired spectral degeneracies.

\subsection{Related Works}
This discussion would be remiss not to explore previous studies that have attempted to fabricate and measure silicon photonic waveguides operating around high-order exceptional point degeneracies. While various research groups have fabricated devices designed to support EPDs in microstrips and at low frequencies (see  Refs.~\cite{nada_frozen_2021,mumcu_partially_2009,jae-youngchungNonInvasiveMetamaterialCharacterization2009,woldeyohannesInternalFieldDistribution2011,haCavityModeControl2010a,othman_experimental_2017,chabanovStronglyResonantTransmission2008}), we restrict our discussion here to optical frequencies and in particular to integrated photonic waveguides (as opposed to photonic crystal waveguides, bound states in the continuum, etc., which have been known to exhibit properties associated with EPDs). As we have demonstrated here, a single device or measurement is insufficient to establish, and sometimes even suggest, operation in or around the FMR. Even a significant spiking of group delay in a device over a narrow frequency range or good agreement between spectral response modeling predictions and measured data is only part of the needed evidence to make such a claim. Both of these observations are important, yet perhaps more important are the \textit{trends} of these observations across identical devices and variations in device length.

EPDs, including the SIP, have anomalous scaling of the group delay with device length. As described in  previous works (see Refs.~\cite{figotin_slow_2011,gutman_slow_2012,figotin_slow_2006} for a few discussions), the maximum delay at the SIP resonance increases proportional to the third power of device length, generally written as $\tau_g \propto N^3$. As has also been described in multiple previous works, fabrication disorder and the sensitivity of periodic structures complicate analysis of measured data. To more reasonably claim a device or set of devices show properties associated with slow-light behavior and the FMR, it is helpful to demonstrate or suggest, in part, this scaling trend. For highly resonant structures and very long devices, compounding disorder imparted by the imperfect fabrication process likely complicated this analysis.

In a previous study, the authors of Ref.~\cite{paulExperimentalDemonstrationFrozenMode2024} fabricate and measure a three-path periodic waveguide that is designed to exhibit the SIP in its eigenmodal dispersion diagram, claiming to have achieved experimental demonstration of the frozen mode. While an important achievement if proven, the authors of the referenced study present a limited analysis of measured data. The data is given at a seemingly coarse measurement wavelength steps (recorded every $\approx 2 \;\mathrm{nm}$, compared to picometer-scale wavelength steps in this work). At this resolution, measurements may miss key spectral features. Furthermore, the authors of Ref.~\cite{paulExperimentalDemonstrationFrozenMode2024} present data for a single device and do not study the group delay in any regard. For  their true time delay device, an analysis of the measured delay would be invaluable in ascertaining the device's properties as has been demonstrably effective in the work presented here. Understanding how the delay changes for different device lengths is important for analyzing properties of SIP-based devices. An additional tool to investigate changes in the analytic or simulation modeling based on measured data is the coalescence parameter. Using the group delay and the coalescence parameter as guides to understand SIP-waveguide behavior provides a more convincing analysis. Because of the presented reasons, we are skeptical of the author's claims in Ref.~\cite{paulExperimentalDemonstrationFrozenMode2024} pertaining to demonstrating an SIP or frozen-mode in a silicon photonics platform.

Another experimental study investigating a fourth-order EPD is Ref.~\cite{burrExperimentalVerificationDegenerate2016}. In that work, the authors fabricate single-path resonators with periodic holes designed to exhibit the DBE in its eigenmodal dispersion diagram. The authors, similar to the data presented here, measured the group delay and further fit Fano resonance lineshapes to measured experimental results. The authors in Ref.~\cite{burrExperimentalVerificationDegenerate2016} compare the quality factor (which is linearly proportional to the group delay) of the first few spectral resonances when varying the number of periods of simulated and fabricated devices. The first resonance closest to the DBE wavelength, generally called the DBE resonance and analogous to the SIP resonance discussed in this work, exhibits the characteristic $N^5$ quality factor scaling trend associated with the DBE. These trends are established both for multiple resonances around the DBE frequency along with multiple device lengths. Based on the author's presented findings and analysis, and in particular based on their analysis of the group delay, they make a strong case for observing the DBE-associated FMR and slow-light behavior in silicon photonics.

In part based on the two references discussed, we do not claim to be the first to demonstrate the FMR or slow-light devices in silicon photonics. We do, however, claim to be the first to provide sufficient evidence suggesting observation of the SIP-associated FMR for longer devices (higher number of unit cells). The measured data for the devices studied here suggest operation in or around the SIP-associated FMR. The remarkable agreement between high-resolution spectral response measurements and corrected model-predicted curves enhances the observation of high group delay peaks near spectral resonances that scale with device length. Additionally, this is the first demonstration in a silicon photonics foundry showing a level of robustness across $300;\mathrm{mm}$ wafer processing.

\section{Conclusion}
\label{ch:Conclusion}
We have presented modeled and measured transfer function and group delay data for a periodic silicon photonic waveguide designed to exhibit the stationary inflection point in its unit cell geometry dispersion relation. We investigated how changing boundary conditions impact our ability to observe properties associated with this eigenmodal degeneracy and how perturbations in the analytic model better match measured results. These perturbations are in-line with variations chip to chip and correspond well to full-wave simulation comparisons with measured data. Around transfer function peaks associated with the stationary inflection point resonances, we observe significant increases in the measured group delay. These trends are consistent across several chips. Calculating the group delay per length, we note how longer structures may impact the ability to observe the expected anomalous scaling of group delay, however we see notable agreement between the delay per length for the perturbed model compared to measured results. The presented and analyzed data indicate the fabricated devices are operating in or near the frozen mode regime around the stationary inflection point.

\section*{Acknowledgment}
This material is based upon work supported by the Air Force Office of Scientific Research award numbers LRIR 24RYCOR008 and FA8655-20-1-7052. N. Furman acknowledges partial support from NSF-AFRL Internship Award No. ECCS-2030029. This material was produced without the use of any generative AI software. The authors are thankful to DS SIMULIA for providing CST Studio Suite.

\section*{Data availability} Data underlying the results presented in this paper are not publicly available at this time but may be obtained from the authors upon reasonable request.

\section*{Disclosures} The authors declare no conflicts of interest.

\bibliographystyle{ieeetr}
\bibliography{main}
\end{document}